\documentstyle[12pt,aasms4]{article}
\begin{document}
\title{The Chemical Compositions of the SRd Variable Stars.  III. 
KK Aquilae, AG Aurigae, Z Aurigae, W Leo Minoris, and WW Tauri}
\author{
Sunetra Giridhar}
\affil{Indian Institute of Astrophysics;
Bangalore,  560034 India\\
giridhar@iiap.ernet.in}

\author{David L.\ Lambert }
\affil{Department of Astronomy; University of
Texas; Austin, TX 78712-1083
\\ dll@astro.as.utexas.edu}

\author{Guillermo Gonzalez}
\affil{Department of Astronomy;
University of Washington;
Seattle, WA 98195-1580\\gonzalez@astro.washington.edu}

\begin{abstract}

Chemical compositions are derived from high-resolution spectra for five
 field SRd variables. 
These supergiants  not previously analysed are shown to be metal-poor:
KK\,Aql with [Fe/H] = -1.2, AG\,Aur  with
[Fe/H] = -1.8, Z\,Aur  with [Fe/H] = -1.4, W\,LMi with [Fe/H] = -1.1, and
WW\,Tau with [Fe/H] =  -1.1.  
Their compositions are, except for two anomalies,
identical to within the measurement errors with
the compositions of subdwarfs, subgiants, and less evolved
giants  of the same [Fe/H]. One anomaly is an  $s$-process enrichment
for KK\,Aql, the first such enrichment reported for a SRd variable.
The second and more remarkable anomaly is
a strong  lithium enrichment for W\,LMi, also a first for field SRds.
The Li\,{\sc i} 6707\AA\ profile
is not simply that of a photospheric line but includes strong absorption from
red-shifted gas, suggesting, perhaps, that lithium enrichment results from
accretion of Li-rich gas.
 This potential  clue to lithium enrichment is discussed in light of
various proposals for lithium synthesis in evolved stars.

{\it Subject headings: stars:abundances -- stars:variables:other (SRd) --
stars:AGB and post-AGB}

\end{abstract}

\section{Introduction}

This series of papers  presents and discusses determinations of the
chemical compositions of the SRd variables. 
Of this mixed bag, we are analysing the Population II members. 
In our first paper (Giridhar, Lambert, \& Gonzalez 1998a\markcite{Gir98a}, 
Paper I), we discussed  four stars:
XY\,Aqr, RX\,Cep, AB\,Leo, and SV\,UMa of which AB\,Leo and SV\,UMa 
were demonstrated to have a low metal abundance ([Fe/H]
$\simeq -1.5$).
% The other two stars were of solar metallicity.
A second paper (Giridhar, Lambert, \& Gonzalez 1999, Paper
II)\markcite{Gir99} 
analysed three metal-poor stars - WY\,And, VW\,Eri, and UW\,Lib. 
Here, we discuss  five  Population II variables (Table\,1)
assigned by Joy (1952\markcite{Joy52}) to the halo population on the basis
of 
radial velocity measurements.

Our spectroscopic analyses are the first for these stars. 
 The elemental abundances
are compared with published results for local metal-poor dwarfs and giants
at an evolutionary stage less extreme than that of our SRd variables. 
Local field dwarfs and giants with very
few exceptions show a common [X/Fe] at a given [Fe/H]. For many
elements [X/Fe] is independent of [Fe/H] over the [Fe/H] range of our
stars.
Sources for the expected [X/Fe] are as follows:
Barbuy (1988\markcite{Bar88}), Carretta, Gratton,
 \& Sneden (2000\markcite{Car00}) and Nissen, Primas, \& Asplund
(2000\markcite{Nis00}) from the [O\,{\sc i}] lines for
O;
 Pilachowski, Sneden, \& Kraft (1996\markcite{Pil96})
 for Na; Gratton, \& Sneden (1988\markcite{Gra88})
for Mg; McWilliam (1997\markcite{McW97}) for Al and Eu; Gratton, \& Sneden
(1991\markcite{Gra91}) for Si, Ca, Sc, Ti, V, Cr, Mn, Co, and Ni;
Gratton \& Sneden (1987)\markcite{Gra87} and Chen et al. 
(2000)\markcite{Che00} for K;
Gratton (1989\markcite{Gra89}) for Mn; Sneden, Gratton, \& Crocker
(1991\markcite{Sne91}) for Cu and Zn; Zhao, \& Magain
(1991)\markcite{Zha91} and Gratton, \& Sneden 
(1994\markcite{Gra94}) for Y, La, Ce, Pr, Nd, and Sm.
Many of these references review previous literature on the
elemental abundances and note the generally close agreement
between the referenced results and other results. For many
elements, the expected value of [X/Fe] should be accurate to
about $\pm$ 0.1 dex. The lack of scatter in [X/Fe] 
for samples composed of stars now in the solar neighborhood
but originating from quite different parts of the Galaxy suggests
that a SRd, if its atmosphere is unaffected by evolutionary effects,
will have the [X/Fe] derived for unevolved stars. 
A `cosmic' scatter in [X/Fe] is seen but only for stars of a lower
metallicity, say [Fe/H] $\leq -2.5$.

\section{Observations and Abundance Analyses}

Spectra were obtained at the McDonald Observatory with the 2.7m telescope
and 
the {\it 2dcoud\'{e}} echelle spectrograph (Tull et al.
1995\markcite{Tul95}). 
In general, each star was observed on more than one occasion and the
spectrum least contaminated by TiO bands was selected for analysis.
Spectra were reduced and analysed by procedures described in Paper I.

As in previous papers in this series, we employed the 1997 version of 
the LTE abundance code, MOOG, developed by Sneden (1973\markcite{Sne73}).
  The model 
atmospheres are from Kurucz (1993) and the MARCS program (Gustafsson et 
al. 1975); the two sets of models give very similar solutions.
Atmospheric 
parameters (Table 2) were derived from Fe\,{\sc i} and Fe\,{\sc ii} 
equivalent widths (EW's).  The effective temperature, T$_{eff}$, is set by 
minimizing the correlation between the Fe\,{\sc i} line abundances and
their 
lower excitation potentials, $\chi_{l}$.  The surface gravity, $g$, is 
determined by forcing ionization equilibrium between Fe\,{\sc i} and
Fe\,{\sc 
ii}.  We also used Ti\,{\sc i}, Ti\,{\sc ii} and Cr\,{\sc i}, Cr\,{\sc ii}
to 
cross-check the derived T$_{eff}$ - $\log g$ pair for each spectrum.  The 
microtubulence velocity parameter, $\xi_{t}$, is set by minimizing the 
correlation between the Fe I abundances and their EW values.  Care has
been 
taken to avoid regions affected by molecular bands.  However, being metal 
poor, the stars in the present study have relatively uncrowded spectra, 
allowing us to measure weak unblended lines. We estimate the measured EW's
to 
be accurate to within 5 to 10 \%.  The uncertainties in T$_{eff}$, $\log g$, and 
$\xi_{t}$ are typically $\pm120$ K, 0.2 dex,
 and $\pm0.2$ km~s$^{-1}$, respectively. 

Our derived abundances are given in Tables  3 -- 7 
as [X/H] = $\log \epsilon$(X/H) $ - \log \epsilon_{\odot}$(X/H) and
[X/Fe], where the H abundance is on the customary scale, and solar 
abundances are taken from Grevesse, Noels, \& Sauval
(1996)\markcite{Gre96}.  
The tables gives the standard error of the mean abundance, as calculated
from 
the line-to-line spread of the abundances. The total error in the absolute 
abundance of a well observed element is generally about $\pm 0.2$ dex when the 
various sources of error (equivalent width, effective temperature, etc.) 
are considered. This does not include systematic errors arising, for
example, 
from the adoption of LTE. In the comparison with the expected abundances 
(Table 8), we cite mean values of  [X/Fe] to $\pm$0.1 dex.

\section{Results and Discussion}

\subsection{KK Aquilae}

For this SRd, which is assigned a period of 88.7 days in the GCVS
(Kholopov et al. 1985), our radial velocity of -243.2 $\pm$ 1.5 km
s$^{-1}$ 
is in good agreement with Joy's (1952) mean velocity of -252 km s$^{-1}$ 
from 7 observations spanning a range of 50 km s$^{-1}$.  Our analysis
shows 
KK\,Aql to have [Fe/H] = -1.2 (Table 2). Elemental abundances are listed
in 
Table 3. The relative abundances [X/Fe] are equal to the expected values
to 
within the errors of measurement for almost all of the elements lighter 
than Y.  An apparent anomaly is Ca for which [Ca/Fe]  = 0.3 is expected 
but -0.2 is derived. A similar result was found from our analyses of cool
RV 
Tauri variables (Giridhar, Lambert, \& Gonzalez 2000).  Potassium  has an 
apparently high abundance: [K/Fe] = 0.8 but about 0.4 is expected. An 
important  signature is the modest overabundance of the heavy elements:
the 
mean [X/Fe] from Y, Zr, Ce, Pr, Ng, and Sm is 0.6 dex against an average 
expected of 0.0 dex. These elements to varying degrees are dominated by
the 
$s$-process (at solar metallicity). Europium, primarily a product of the 
$r$-process, has [Eu/Fe] = 0.5, but this is effectively the expected
value.  
KK\,Aql is the first SRd in our sample to show evidence of $s$-process 
enrichment, which may be presumed to have resulted from operation of the 
$s$-process in the He-shell of an AGB star.  This He-shell may belong to
the 
SRd itself or to a companion star that transferred mass to the SRd,
presumably 
at an earlier stage of evolution.

\subsection{AG Aurigae}

This SRd was observed at a phase when the TiO bands were strong. The
radial 
velocity of -193 $\pm 1.4$ km s$^{-1}$ from atomic lines
matches perfectly the mean velocity reported by Joy (1952) from
15 observations spanning 25 km s$^{-1}$. The Fe abundance corresponds to
[Fe/H] = -1.8. Relative to the expected [X/Fe], there appear to
be some  anomalies. The most striking are
[Mg/Fe] = -0.4 from a single line, and [Ca/Fe] = -0.3  for which the expected
values for   both are  0.3.
Other  light elements -- Si, Ti, V, Cr, Mn, Co, Ni, and Cu --
 provide an [X/Fe] that is the expected value to
within the errors of measurement.
The heavy elements    --- Y to Sm --- show a
 scatter in [X/Fe], but
 the mean from the 7 elements is close  to
expectation with just a hint of $s$-process enrichment.
 The scatter presumably reflects the difficulty of determining
abundances based on measuring just a few
lines in a crowded spectrum.

\subsection{Z Aurigae}

Z Aurigae is variously classified as a Mira and a SRd variable. Two
studies
of the AAVSO magnitude estimates have shown that the period has changed
(Lacy 1973; Percy, \& Colivas 1999). Percy \& Colivas remark that the 
period changed  abruptly  around 1929 (JD 2425575). 
  Our measured radial velocity of
-174 $\pm$ 1.3 km s$^{-1}$ is consistent with Joy's (1952)
 mean velocity of -165 km s$^{-1}$ from 9 observations
covering 35 km s$^{-1}$.
The star has [Fe/H] = -1.4 with abundances [X/Fe] close to the
expected values.  Titanium with [Ti/Fe] = -0.1 and V with [V/Fe] = -0.3 
may be slightly anomalous; [Ti/Fe] = 0.3 and [V/Fe] = 0.0 are expected. 
Sulphur may be slightly overabundant but the result is based on
a single line.  Europium shows the expected small enrichment relative to
Fe. Other heavy elements have their expected abundances: [X/Fe]
 = 0.0 from five elements; Z Aur is not $s$-process enriched.

\subsection{W Leo Minoris}

The abundance analysis is based on the 1999 February 2 spectrum taken
at a time when the TiO bands were very weak;
the spectrum is that of an early-mid K giant. Other spectra
containing  TiO bands were examined - see below. A remarkable
feature is a strong Li\,{\sc i} 6707\AA\ line with a complex profile
 (Figure 1). 
This star does not have such
an extreme radial velocity as the other stars: we find +51.0 $\pm$ 1.2
km s$^{-1}$ which compares  with Joy's (1952) mean velocity of
60 km s$^{-1}$ from 10 observations
over a range of 55 km s$^{-1}$.
At [Fe/H] = -1.1, W LMi is one of   two least metal-poor of the five
stars.
The relative abundances [X/Fe] (Table 6)
 are all close to the expected values for
a halo star; lithium is discussed below.
 Europium with [Eu/Fe] = 0.9 appears
overabundant by about 0.5 dex, but this result is based  
on a single line. 

The remarkable feature of W\,LMi's spectrum is a deep broad absorption
feature at 6707\AA. This is naturally identified as the Li\,{\sc i}
resonance doublet. Our spectrum synthesis (Figure 1) clearly shows that
absorption is not simply due to photospheric lithium, assumed to be $^7$Li.
Addition of $^6$Li
does not materially alter
the predicted line profile.
 The abundance adopted for the `best' fit is
log$\epsilon$(Li) $\simeq$ 1.9. There is clearly additional absorption 
present. Although the extra absorption is mainly to the red of the
photospheric line, there is a  mismatch between the observed and
synthetic profiles in the latter's blue wing; one might question the
degree
to which photospheric Li is at all a contributor.  The observed profile
suggests that weak emission may be filling in a single broad line.
The excited Li\,{\sc i} 6103\AA\ line which
  is present (at the edge of an order)
as a single absorption line with a photospheric profile gives a much
higher abundance log$\epsilon$(Li) $\simeq$ 3.2; the line
does not have either the red component or the extended blue wing of the
resonance line. (The excited Li\,{\sc i}
line at 8126\AA\ was not recorded by our spectrum.) Judged by results of
non-LTE calculations for giants of a somewhat higher surface gravity
(e.g., Balachandran et al. 2000\markcite{Bal00}),
an LTE analysis of the 6103\AA\ line should return a slightly lower not a
higher Li abundance than the 6707\AA\ line. We suppose that the
discrepancy here may be due to either a filling in of the 6707\AA\ 
photospheric absorption line by emission or to the failure of the
star's upper atmosphere to track the theoretical structure used in the
synthesis.
It seems  obvious that lithium must be present
in great abundance.  The discussion that follows is not
greatly dependent on identification of log$\epsilon$(Li) $\sim 2$ or  3 as the
stellar abundance. 

The key issue is that lithium has been added to W LMi's atmosphere
since its birth.
W LMi's initial Li abundance would have been close
to that of the Spite
plateau, log$\epsilon$(Li) $\simeq 2.2$, but 
in evolution from the
main sequence to the red giant branch, atmospheric lithium would have
been greatly diluted, and, in addition, 
lithium may have been destroyed on the main sequence. In short, the
surface lithium 
 has been  replenished, presumably after the initial dilution as a red
giant. 
%Presence of circumstellar lithium suggests that lithium replenishment
%was accompanied by accretion and possibly  ejection of material. 

Recent observations have shown that lithium enrichment occurs, albeit
rarely,
in  highly evolved low mass metal-poor O-rich
 stars. Carney, Fry, \& Gonzalez  (1998)\markcite{Car98} found lithium
enrichment in the RV Tauri-like variable V42  in the globular
cluster M5.  Kraft et al. (1999)\markcite{Kra99}
  and  Smith, Shetrone, \& Keane (1999)\markcite{Smi99}
each discovered a  Li-rich red giant in a globular cluster.
Hill \& Pasquini (1999)\markcite{Hil99}
 found a Li-rich giant at a luminosity higher than
that of He-core burning clump stars in an open cluster of metallicity
[Fe/H] = -0.5. 
  We presume
that the same unknown mechanism of lithium enrichment
that  operated in these  cluster stars was at work in W LMi.

 Our spectra
provide a  clue to that mechanism - the Li\,{\sc i} 6707\AA\ line
and other resonance and strong low excitation lines are double with
a strong red-shifted absorption component.
In the 1999 February 2 spectrum,  the Na D, K\,{\sc i} 7699\AA, and the
Ba\,{\sc ii} 4554\AA\ resonance lines show a double absorption profile
similar to that of the Li\,{\sc i} 6707\AA\ line in Figure 1. The
separation of the two components is consistent with the 21.5 km s$^{-1}$
measured for the 6707\AA\ line. The Na D lines also show blue-shifted
components at -64 and -89 km s$^{-1}$, which are 
presumably  interstellar lines.
Strong low excitation lines also show 
the red component, as in the Mg\,{\sc i} 8807\AA, and the Ba\,{\sc ii}
lines at 5853, 6141, and 6497\AA. 
A similar broad, likely double, profile was observed by Smith et al. for
the Li\,{\sc i} 6707\AA\ line, as well as the  K\,{\sc i} 7699\AA\ line,
in the Li-rich giant in the globular cluster NGC362. This cluster
star, a SRd  with a period of about 100 days, is similar to W LMi being
about 400K cooler but of similar metallicity and surface gravity.

Examination of other spectra of W LMi taken at times when the TiO bands
were stronger shows that the line doubling is probably present at all
phases (Figure 2). The line separation remains about constant, i.e., the
red component shifts with the pulsational velocity. This component of
the Li\,{\sc i} 6707\AA\ line
presumably originates in the upper atmosphere in cool gas. One
possibility is that we are detecting infalling gas from a circumstellar
shell or gas ejected at an earlier pulsation cycle. The velocity
of infall of about 20 km s$^{-1}$ is approximately half the expected 
free fall velocity.  Alternatively, the
doubling of the absorption lines may signify the presence of a shock in
the
upper atmosphere with the shock separating cooler and hotter gas. If the
latter scenario is correct, we would expect to see a stronger phase
dependence
of the strength and separation of the blue and red components. 
Such prominent line doubling is not  unique among SRds to W LMi.
Resonance lines, especially the
Na D lines, are doubled in Z Aur, AB Leo, and  WW Tau.
The Na D
splitting for AB Leo and Z Aur is about 15 km s$^{-1}$.
Z Aur's Na D components near zero
heliocentric velocity are plausibly attributed to  interstellar sodium but
appear also  split  by about 15 km s$^{-1}$, a presumably fortuitous
similarity with the stellar line splitting. 

Lithium-enriched giants of
near-solar metallicity are known and several proposals for enriching
a stellar atmosphere in lithium have been offered (see review by 
Charbonnel \& Balachandran 2000\markcite{Cha00}). All but one of the
proposals
invokes  conversion of $^3$He to $^7$Li in the interior of the
Li-rich star; the exception is 
lithium enrichment resulting from accretion of a planet,
an idea first advocated by Alexander (1967)\markcite{Ale67}.
Accretion of a terrestrial planet will enhance the lithium abundance
to levels possibly in excess of the star's initial abundance. Early
accretion of the planet is likely  followed by its destruction
at the base of the convective envelope; lithium
enrichment in this (and most other) scenarios is a passing phase.
Terrestrial
planets might be expected to  orbit close to a star and so be
 accreted early in a giant's
life with the possibility that lithium is destroyed subsequently.
 Therefore, the planet scenario seems ill-suited as an explanation
for
W\,LMi, a highly evolved red giant. Moreover, terrestrial planets are
unlikely
to form around metal-poor stars. Accretion of a brown dwarf will not
greatly enhance the stellar lithium abundance. 
(A simpler process of mass transfer
from a Li-rich AGB companion star seems improbable too. W LMi does not
presently have such an AGB companion or a  luminous post-AGB derivative.
We cannot exclude a white dwarf descendant but, as accretion of Li-rich
material may be continuing, it seems to be necessary to invoke 
long-term storage of the companion's ejecta.)

Production, if it is the key to lithium
replenishment for W LMi,
 was most probably by the Cameron-Fowler (1971)\markcite{Cam71}
 mechanism involving
the reaction sequence $^3$He($^4$He,$\gamma$)$^7$Be(e$^-$,$\nu_e$)$^7$Li
initiated at high temperatures but
which to be efficient must occur in a convective or explosive layer such
that the $^7$Be and $^7$Li are transported rapidly to cool layers and
avoid losses due to proton-capture. The $^3$He is a fruit of incomplete
$pp$-chain processing of hydrogen. Since the opening reaction of the
$pp$-chain
is controlled by the weak interaction and is, therefore, very slow, the
participating $^3$He nuclei must be survivors of production during
the main sequence.

 Theory predicts that a   very luminous AGB star
experiences
H-burning at the base of its convective envelope and may enrich the
surface in 
Li via the Cameron-Fowler mechanism
(Scalo 1976\markcite{Sca76};
 Sackmann, \& Boothroyd 1992\markcite{Sca92}). In the case of
metal-poor stars, Lattanzio, Forestini, \& Charbonnel
 (2000)\markcite{Lat00} show that
the surface ratio of C/O stays below unity for a considerable fraction of
the star's life on the AGB;  loss of C to N to H-burning via the CN-cycle 
is offset by the dredge-up of freshly synthesized $^{12}$C in the deeper
He-burning shell.
These calculations are fully consistent with respect to luminosity
and lithium abundance with the observations of Li-rich
S stars in the Magellanic Clouds
(Smith \& Lambert 1989\markcite{Smi89}, 1990\markcite{Sci90}, Smith et al.
1995\markcite{Smi95}).

W\,LMi is obviously O-rich because the spectrum  at the appropriate phases
is
cluttered with TiO bands and not  C$_2$ bands. This tidy explanation of
W LMi's lithium
has two
serious 
drawbacks. Present calculations predict Li synthesis to occur only in
intermediate
mass AGB stars, say $M \sim 4M_\odot$, although calculations for
metal-poor
AGB stars have not yet been extended to low mass. Nonetheless,
it  seems 
 likely that a metal-poor
star like W\,LMi is of  low mass, say $M \sim 1M_\odot$, will not
develop the necessary hot bottomed convective envelope. 
In addition, the Cameron-Fowler mechanism is not triggered
 until a star has experienced many He-shell flashes
and subsequent third dredge-up of carbon and $s$-process elements; the
Magellanic Cloud Li-rich giants are recognized as S stars.
W\,LMi, however, is not enriched in the $s$-process elements.

Then, one must suppose that the conversion of $^3$He to $^7$Li occurs
at a different (and earlier) stage of evolution. One is tempted given
the location of the Li-rich star in NGC362 to invoke the helium-core
flash as the trigger for mixing and production.
 Of the published conversion mechanisms (mostly, speculative musings),
 only one imagines mass loss to accompany or closely follow lithium
production.
 de la Reza et al. (1996\markcite{del66},
 1997\markcite{del97}) suggested that
lithium production was associated with mass ejection; a correlation
with infrared excesses arising from circumstellar
dust promoted this suggestion. If the infalling gas is associated
with an extended envelope or circumstellar shell, W LMi suggests a
link between a  lithium-rich
atmosphere and  an external gas reservoir that is also Li-rich.
The link, however, may not be directly as imagined by de la Reza
and his colleagues because the solar-metallicity Li-rich giants are,
as shown by Charbonnel \& Balachandran (2000\markcite{Cha00}), all found
at
a similar luminosity but  below the luminosity of W LMi.
 Given that the luminosity of these Li-rich giants is
rather tightly defined, it would seem that lithium
enrichment is a transitory phase.

\subsection{WW Tauri}

Our measured radial velocity of -114.0 $\pm$ 1.1 km s$^{-1}$ is in good
agreement with  the -110 km s$^{-1}$ mean velocity obtained by
Joy (1952)\markcite{Joy52} from 6 observations over a range of 20 km
s$^{-1}$.

WW Tau has a normal composition for a [Fe/H] = -1.1 star. Calcium appears
slightly underabundant at [Ca/Fe] =  0.0, but a slight apparent
underabundance of Ca is common among SRd
variables and may indicate a systematic error such as non-LTE effects.
Scandium may be slightly overabundant. The heavy elements do not show
evidence of enrichment: the mean abundance from five elements each
represented by two or more lines is [X/Fe] = 0.2.

\section{Concluding Remarks}

Table 8 shows the mean observed and the expected [X/Fe] for the five
stars.
In compiling  this table,
 we have included only elements measured in 4 or 5 stars of this
paper, and for the heavy elements we omitted KK Aql ($s$-process enriched)
and included elements measured in 3 or 4  stars. 
 For the `metals' from Na through the
iron-group to Cu and Zn, the stars have the relative abundances expected
from
studies of metal-poor dwarfs and less evolved giants.
In particular, we note the underabundance of Mn and Cu that is seen also
in
normal stars; [X/Fe] for Mn and Cu for normal
stars  are slightly dependent on [Fe/H].
(Luck \& Bond (1985\markcite{Luc85})
 found a remarkable Mn deficiency for TY\,Vir.)
 We earlier alluded to the
slight discrepany between the observed and expected value of [Ca/Fe], and
to the similar discrepancy found for cool RV Tauri variables.
Adjustments to the atmospheric parameters made to eliminate the Ca
underabundance create additional anomalies. An alternative
 contributing factor may
be non-LTE effects on the Ca\,{\sc i} lines. Drake (1991\markcite{Dra91})
predicted a nearly 0.2 dex increase in the Ca abundance when non-LTE
effects were included for a model atmosphere at T$_{\rm eff}$ = 4500 K
and $\log g$ = 1. We consider it quite unlikely that the Ca underabundance
can be taken as indicating that these stars are descendants of the
$\alpha$-poor subdwarfs discovered by Nissen \& Schuster (1987\markcite{Nis97}).  

These five
 stars like the metal-poor SRds analysed in Papers I and II are on the AGB
or near the tip of the RGB, and as such may be the immediate
progenitors of those  RV\,Tauri variables that are intrinsically
 metal-poor.
What seems remarkable
 is the widespread
lack of an $s$-process enrichment. A similar surprise was offered by the
metal-poor RV Tauri variables (Giridhar et al. 2000\markcite{Gir00}).
 Only KK\,Aql is clearly enriched out of 
the sample of 11 SRd stars analysed by us. A couple of stars hint at mild
enrichments ([$s$/Fe] $\simeq 0.2$) but the remainder are
resolutely unenriched.  This result would seem to imply
that SRds evolve off either the RGB or the AGB  before thermal pulses have
enriched the envelope in $s$-process products and carbon from the
He-shell. A thorough analysis of the C, N, and O elemental and
isotopic abundances for SRds, and, in particular, a detailed comparison of
the composition of the $s$-process enriched KK\,Aql
with other SRds should be instructive.

An outstanding result  is the presence of lithium in and/or
near W\,LMi's atmosphere. This is the first example
of lithium enrichment in our survey of SRds.  In common with
the Li-rich star in the  globular cluster NGC362 (Smith et al. 1999)
\markcite{Smi99}, the lithium enrichment
may be associated
with  Li-rich circumstellar gas, suggesting that
 lithium production in the
star was followed by ejection of material that is now returning to the
star. One hopes that completion of our survey of SRds will reveal
additional Li-rich examples (V Pyx is just such a case),
 and fresh insights into lithium enrichment. Two extreme scenarios may
be envisaged: Is
the rarity of observed lithium enrichment the result of effective
lithium production occurring in very few stars or  does
production occur in all stars to be followed by  rapid destruction? 

We thank John Lattanzio, and George Wallerstein for helpful comments, and
Suchitra Balachadran for use of her 6103\AA\ line list.
This research has been supported in part by the Robert A.\, Welch
Foundation of
Houston, Texas and the National Science Foundation (grant AST-9618414).

\clearpage
\clearpage
\section*{Figure Caption}
\figcaption{The spectrum of W LMi around 6707\AA. The dotted line is the
observed spectrum from 1999 February 2. The solid line is a synthesis
for a lithium abundance log$\epsilon$(Li) = 1.9. Three Fe\,{\sc i}
lines are identified. Note the mismatch between the predicted photospheric
Li\,{\sc i} profile and the broad double line.}
\figcaption{The spectrum of W LMi around 6707\AA\ on 1998 January 24,
1999 February 2, and 2000 May 13. The strong Li\,{\sc i} 6707\AA\
line is present in all spectra as an apparent blend of two lines separated
by about 20 km s$^{-1}$.}
\end{document}

\begin{deluxetable}{lccr}
\tablewidth{0pt}
\tablecaption{The Program Stars}
\tablehead{
\colhead{Star} & \colhead{V} 
& \colhead{Sp. Type} & \colhead{Period}   
}   
\startdata
 KK Aql & 11.5 -12.8   & G2e - G6  & 88.7  \nl
 AG Aur & 10.00 - 13.1  & G2e - K0  & 96.0  \nl
 Z Aur  & 9.20 - 11.7 &  G0e - G6    & 134.8  \nl
  W LMi & 10.50 - 13.5 &  G2e - K2 &117.2  \nl
 WW Tau & 9.00 -12.9 & G4e - K2 & 116.4 \nl
\enddata
\end{deluxetable}

\clearpage

\begin{deluxetable}{llrlcrlr}
\tablewidth{7in}
\tablecaption{Stellar Parameters Derived from the
Fe-line Analyses}
\tablehead{
\colhead{Star} &  \colhead{Model\tablenotemark{a}} &
 \colhead{$\xi_{\rm t}$} & \multicolumn{2}{c}{Fe\,{\sc i}\tablenotemark{b}} 
& & \multicolumn{2}{c}{Fe\,{\sc ii}}  \\ \cline{2-2} \cline{4-5} \cline{7-8}
\colhead{~~UT Date} &\colhead{T$_{\rm eff}$, $\log g$,
[Fe/H]} & \colhead{(km~s$^{-1}$)} & \colhead{$\log \epsilon$} &
\colhead{n} & &\colhead{$\log \epsilon$} & \colhead{n} 
}
\startdata
KK Aql  & & & &&&  &  \nl
~~1999 Nov 3  & 4300, 0.5, $~~-1.2$ & 2.5\phantom{000} & $6.28 \pm 0.20$ &
36 && $6.21 \pm 0.18$ & 2   \nl
AG Aur  & & & &&&  &  \nl
~~1998 Jan 25 & 4000, 0.0, $~~-1.7$ & 3.2\phantom{000} & $5.75 \pm 0.17$ &
35 && $5.76 \pm 0.20$ & 2   \nl
 Z Aur & & &&  & & & \nl
~~1998 Jan 25& 4300, 0.0, $~~-1.4$ & 2.5\phantom{000} & $6.12 \pm 0.16$ &
58 && $6.07 \pm 0.13$ &  11   \nl
  W LMi & & && & & &  \nl
~~1999 Feb 2 & 4300, 0.8, $~~-1.1$ & 2.3\phantom{000} & $6.45 \pm 0.16$ &
100 && $6.41 \pm 0.14$ &12   \nl
 WW Tau & & && & & &  \nl
~~1998 Jan 25 & 4300, 0.8, $~~-1.1$ & 2.7\phantom{000} & $6.38 \pm 0.17$ &
 98 && $6.35 \pm 0.20$ & 5   \nl
\enddata
\tablenotetext{a}{$T_{\rm eff}$ in K, log $g$ in cgs, [Fe/H] in dex.}
\tablenotetext{b}{$\log \epsilon$ is the mean abundance relative to H (with 
$\log N_{\rm H} = 12.00$). The solar value of $\log \epsilon$(Fe) is 7.51. The 
standard deviations of the means, as calculated from the line-to-line scatter, 
are given. $n$ is the number of lines used in the analysis.}

\end{deluxetable}

\clearpage

\begin{deluxetable}{lclrc}
\tablecaption{Elemental Abundances for KK Aql}
\tablewidth{4in}
\tablehead{
\colhead{Species} & \colhead{ $\log \epsilon_{\odot}$\tablenotemark{a}} &
\colhead{[X/H]} & \colhead{$n$\tablenotemark{b}} &
\colhead{[X/Fe]}
}
\startdata
O\,{\sc i}& 8.87 &$-0.44$$\pm$0.15&2&$+0.79$\nl
Si\,{\sc i}& 7.55 &$-0.90$$\pm$0.36&3&$+0.33$\nl
K\,{\sc i} & 5.12 &$-0.38$          & 1  &$+0.85$\nl
Ca\,{\sc i} & 6.35 &$-1.46$$\pm$0.15  &3 & $-0.23$\nl
Ti\,{\sc i}&4.98 & $-0.83$$\pm$0.20  &19 & $+0.40$\nl
Ti\,{\sc ii}&4.98 & $-0.93$$\pm$0.24  &  2& $+0.30$\nl
 V\,{\sc i} &4.01 & $-0.95$$\pm$0.17  &10 & $+0.28$\nl
Cr\,{\sc i}&5.67 & $-1.00$$\pm$0.20 & 3 & $+0.23$\nl
Cr\,{\sc ii}&5.67 & $-1.08$& 1 & $+0.15$\nl
Mn\,{\sc i}&5.39 & $-1.05$$\pm$0.27 & 3 & $+0.18$\nl
Fe\,{\sc i}&7.51 & $-1.23$$\pm$0.20  &  36 &\nodata \nl
Fe\,{\sc ii}&7.51 & $-1.29$$\pm$0.18 & 3   &\nodata \nl
Ni\,{\sc i} & 6.25 & $-1.06$$\pm$0.21 &18 &$+0.17$ \nl
Cu\,{\sc i}& 4.25 & $-1.77$  & 1 &$-0.54$ \nl
Y\,{\sc ii}& 2.23 & $-0.99$$\pm$0.20  & 2 &$+0.26$ \nl
Zr\,{\sc i} & 2.60 & $-0.50$$\pm$0.08  & 3 &$+0.73$ \nl
Ce\,{\sc ii}& 1.55 & $-0.40$$\pm$0.16  & 2 &$+0.83$ \nl
Pr\,{\sc i} & 0.71 & $-0.36$$\pm$0.22  & 2 &$+0.87$ \nl
Nd\,{\sc i}& 1.50 & $-0.67$$\pm$0.22 & 7 &$+0.56$ \nl
Sm\,{\sc ii}& 1.00 & $-0.70$$\pm$0.23  &  7 &$+0.53$ \nl
Eu\,{\sc ii}& 0.51 & $-0.74$$\pm$0.23 &  7 &$+0.49$ \nl
\enddata
\tablenotetext{a}{The solar abundances are taken from  Grevesse, Noels
 \& Sauval (1996).}
\tablenotetext{b}{Number of lines.  These comments also apply to the following 
four tables.}
\end{deluxetable}

\clearpage

\begin{deluxetable}{lllrl}
\tablecaption{Elemental Abundances for AG Aur}
\tablewidth{3.7in}
\tablehead{
\colhead{Species}& \colhead{ $\log \epsilon_{\odot}$} &
\colhead{[X/H]}&
\colhead{$n$} & \colhead{[X/Fe]}}
\startdata
Mg\,{\sc i} & 7.58 &$-2.22$& 1  & $-0.38$\nl
Si\,{\sc i} & 7.55 &$-1.33$$\pm0.06$ & 3  &$+0.51$\nl
Ca\,{\sc i} & 6.35 &$-2.15$$\pm$0.12 &3 & $-0.31$\nl
Ti\,{\sc i} &4.98 & $-1.49$$\pm$0.24 &16 & $+0.35$\nl
Ti\,{\sc ii} &4.98 & $-1.40$$\pm$0.20 &  3& $+0.44$\nl
 V\,{\sc i}  &4.01 & $-1.87$$\pm$0.18 & 3& $-0.03$\nl
Cr\,{\sc i}  &5.67 & $-2.18$$\pm$0.30 & 5 & $-0.34$\nl
Mn\,{\sc i}  &5.39 & $-1.99$$\pm$0.27 & 2 & $-0.15$\nl
Fe\,{\sc i}  &7.51 &$-1.83$$\pm$0.15 &  35 &\nodata \nl
Fe\,{\sc ii} &7.51 & $-1.84$$\pm$0.14 & 2   &\nodata \nl
Co\,{\sc i} & 4.91 & $-1.69$$\pm$0.20&5 &$+0.15$ \nl
Ni\,{\sc i} & 6.25 & $-1.80$$\pm$0.17&12 &$+0.04$ \nl
Cu\,{\sc i} & 4.25 & $-2.14$& 1 &$-0.30$ \nl
Y\,{\sc ii} & 2.23 & $-1.79$ & 1 &$+0.04$ \nl
Zr\,{\sc i} & 2.60 & $-1.49$$\pm$0.05  & 4 &$+0.35$ \nl
Zr\,{\sc ii} & 2.60 & $-1.25$ & 1 &$+0.59$ \nl
Ba\,{\sc ii} & 2.13 & $-2.23$$\pm$0.27& 2 &$-0.39$ \nl
Ce\,{\sc ii} & 1.55 & $-1.93$& 1 &$-0.09$ \nl
Nd\,{\sc ii} & 1.50 & $-1.37$$\pm$0.16& 6 &$+0.47$ \nl
Sm\,{\sc ii}& 1.00 & $-1.27$$\pm$0.23&  4 &$+0.57$ \nl
\enddata
\end{deluxetable}

\clearpage

\begin{deluxetable}{lllll}
\tablecaption{Elemental Abundances for Z Aur}
\tablewidth{3.7in}
\tablehead{
\colhead{Species}& \colhead{ $\log \epsilon_{\odot}$} &
\colhead{[X/H]}  &
\colhead{$n$}&\colhead{[X/Fe]}}
\startdata
O\,{\sc i} & 8.87 &$-1.05$  &1 & $+0.36$ \nl
Na\,{\sc i} & 6.32 & $-1.59$$\pm$0.14 &2 &$-0.18$  \nl
Mg\,{\sc i} & 7.58 &$-1.12$$\pm$0.09 & 2  & $+0.30$ \nl
Al\,{\sc i} & 6.47 &$-1.43$$\pm$0.10& 3&$-0.02$\nl
Si\,{\sc i} & 7.55 &$-1.04$$\pm$0.13 & 9  &$+0.38$\nl
S\,{\sc i} & 7.21 & $-0.71$&1 & $+0.70$ \nl
Ca\,{\sc i} & 6.35 &$-1.31$$\pm$0.12 &8 & $+0.10$\nl
Sc\,{\sc i} & 3.13 & $-1.27$$\pm$0.19  &4 & $+0.15$\nl
Ti\,{\sc i} &4.98 & $-1.56$$\pm$0.25 & 8& $-0.15$\nl
Ti\,{\sc ii} &4.98 & $-1.51$$\pm$0.20 &  2& $-0.10$\nl
 V\,{\sc i} &4.01 & $-1.68$$\pm$0.18 & 11& $-0.26$\nl
Cr\,{\sc i} &5.67 & $-1.57$$\pm$0.20 & 9 & $-0.16$\nl
Cr\,{\sc ii} &5.67 & $-1.55$$\pm$0.20 & 4 & $-0.14$\nl
Mn\,{\sc i}  &5.39 & $-1.82$$\pm$0.27 & 6 & $-0.41$\nl
Fe\,{\sc i} &7.51 & $-1.39$$\pm$0.16 &  58 &\nodata \nl
Fe\,{\sc ii} &7.51 & $-1.44$$\pm$0.13 & 11  &\nodata \nl
Co\,{\sc i} & 4.91 & $-1.38$$\pm$0.18&5 &$+0.03$ \nl
Ni\,{\sc i} & 6.25 & $-1.54$$\pm$0.17&15 &$-0.13$ \nl
Cu\,{\sc i} & 4.25 & $-1.47$$\pm$0.08& 2 &$-0.06$ \nl
Zn\,{\sc i} & 4.60 & $-1.34$$\pm$0.07 &2 &$+0.07$ \nl
Y\,{\sc ii} & 2.23 & $-1.42$$\pm$0.13& 4 &$-0.01$ \nl
Ba\,{\sc ii} & 2.13 & $-1.60$& 1 &$-0.19$ \nl
Ce\,{\sc ii} & 1.55 & $-1.51$$\pm$0.30& 4 &$-0.10$ \nl
Pr\,{\sc ii} & 0.75 & $-1.16$& 1 &$+0.25$ \nl
Nd\,{\sc ii} & 1.50 & $-1.29$$\pm$0.16& 10 &$+0.12$ \nl
Sm\,{\sc ii} & 1.00 & $-1.45$$\pm$0.03&  2 &$-0.04$ \nl
Eu\,{\sc ii} & 0.51 & $-0.99$ & 1  &$+0.42$ \nl
\enddata
\end{deluxetable}

\clearpage

\begin{deluxetable}{lllll}
\tablecaption{Elemental Abundances for W LMi}
\tablewidth{4in}
\tablehead{
\colhead{Species}& \colhead{ $\log \epsilon_{\odot}$} &
\colhead{[X/H]}&
\colhead{$n$} &  \colhead{[X/Fe]} }
\startdata
O\,{\sc i} & 8.87 & $-0.50$$\pm$0.05 &3 & $+0.58$ \nl
Na\,{\sc i}& 6.32 & $-1.37$$\pm$0.09 &3 &$-0.29$  \nl
Mg\,{\sc i} & 7.58 &$-0.82$& 1 & $+0.26$ \nl
Si\,{\sc i} & 7.55 &$-0.87$$\pm$0.15 & 7  &$+0.21$\nl
Ca\,{\sc i} & 6.35 &$-0.94$$\pm$0.06 &5 & $+0.15$\nl
Sc\,{\sc i} & 3.13 & $-0.96$&1 & $+0.12$\nl
Sc\,{\sc ii}& 3.13 & $-1.01$$\pm$0.21  &6 & $+0.07$\nl
Ti\,{\sc i} &4.98 & $-1.03$$\pm$0.20 & 25& $+0.05$\nl
Ti\,{\sc ii}&4.98 & $-0.88$$\pm$0.22 &  4& $+0.20$\nl
 V\,{\sc i} &4.01 & $-0.94$$\pm$0.21 & 22& $+0.14$\nl
Cr\,{\sc i} &5.67 & $-1.09$$\pm$0.17 & 7 & $-0.01$\nl
Cr\,{\sc ii}&5.67 & $-1.07$$\pm$0.16 & 5 & $+0.01$\nl
Mn\,{\sc i} &5.39 & $-1.43$$\pm$0.25 & 6 & $-0.35$\nl
Fe\,{\sc i}&7.51 & $-1.06$$\pm$0.16 &  98 &\nodata \nl
Fe\,{\sc ii}&7.51 & $-1.10$$\pm$0.13 & 12  &\nodata \nl
Co\,{\sc i} & 4.91 & $-0.97$$\pm$0.17&3 &$+0.11$ \nl
Ni\,{\sc i} & 6.25 & $-1.19$$\pm$0.18&23 &$-0.11$ \nl
Zn\,{\sc i} & 4.60 & $-1.26$&1 &$-0.18$ \nl
Cu\,{\sc i} & 4.25 & $-1.55$&1&$-0.37$\nl
Y\,{\sc ii}& 2.23 & $-1.06$$\pm$0.22& 4 &$+0.02$ \nl
Zr\,{\sc i} & 2.60 & $-0.99$$\pm$0.19& 5 &$+0.09$ \nl
Mo\,{\sc i} & 1.92 & $-0.62$& 1 &$+0.56$ \nl
La\,{\sc ii}& 1.22 & $-0.94$$\pm$0.20& 3 &$+0.14$ \nl
Ce\,{\sc ii} & 1.55 & $-1.17$$\pm$0.16& 7 &$-0.09$ \nl
Pr\,{\sc ii} & 0.75 & $-0.65$$\pm$0.26& 2 &$+0.43$ \nl
Nd\,{\sc ii} & 1.50 & $-0.80$$\pm$0.20& 9 &$+0.28$ \nl
Sm\,{\sc ii} & 1.00 & $-0.65$$\pm$0.18& 12 &$+0.43$ \nl
Eu\,{\sc ii}& 0.51 & $-0.14$& 1  &$+0.92$ \nl
\enddata
\end{deluxetable}

\clearpage

\begin{deluxetable}{lllll}
\tablecaption{Elemental Abundances for WW Tau}
\tablewidth{4in}
\tablehead{
\colhead{Species}& \colhead{ $\log \epsilon_{\odot}$} &
\colhead{[X/H]}&
\colhead{$n$} & \colhead{[X/Fe]}}
\startdata
O\,{\sc i} & 8.87 & $-0.59$&1&$+0.55$ \nl
Na\,{\sc i} & 6.32 & $-1.19$$\pm$0.04 &2 &$-0.05$  \nl
Mg\,{\sc i}& 7.58 &$-1.01$$\pm$0.0& 2  & $+0.13$ \nl
Al\,{\sc i}& 6.47 &$-0.76$$\pm$0.05& 2  &$+0.38$\nl
Si\,{\sc i}& 7.55 &$-0.69$$\pm$0.20 & 8  &$+0.45$\nl
Ca\,{\sc i} & 6.35 &$-1.16$$\pm$0.18 &9 & $-0.03$\nl
Sc\,{\sc ii}& 3.13 & $-0.84$$\pm$0.18  &5 & $+0.29$\nl
Ti\,{\sc i} &4.98 & $-0.89$& 1& $+0.25$\nl
Ti\,{\sc ii} &4.98 & $-1.04$$\pm$0.12 &  6& $+0.10$\nl
Cr\,{\sc i}&5.67 & $-1.16$$\pm$0.23 & 6 & $-0.02$\nl
Cr\,{\sc ii}&5.67 & $-1.08$$\pm$0.30 & 3 & $+0.06$\nl
Mn\,{\sc i}&5.39 & $-1.47$$\pm$0.02 & 2 & $-0.33$\nl
Fe\,{\sc i} &7.51 & $-1.13$$\pm$0.18 &  98 &\nodata \nl
Fe\,{\sc ii}&7.51 & $-1.16$$\pm$0.21 & 13  &\nodata \nl
Co\,{\sc i} & 4.91 & $-0.88$$\pm$0.03&5 &$+0.26$ \nl
Ni\,{\sc i} & 6.25 & $-1.17$$\pm$0.19&30 &$-0.03$ \nl
Cu\,{\sc i} & 4.25 & $-1.19$$\pm$0.34 &2&$-0.05$\nl
Zn\,{\sc i} & 4.60 & $-1.18$&1 &$-0.04$ \nl
Y\,{\sc ii}& 2.23 & $-1.02$$\pm$0.12& 4 &$+0.12$ \nl
Zr\,{\sc i} & 2.60 & $-1.01$$\pm$0.19& 5 &$+0.13$ \nl
La\,{\sc ii} & 1.22 & $-0.58$& 1 &$+0.56$ \nl
Ce\,{\sc ii}& 1.55 & $-1.05$$\pm$0.09& 3 &$+0.09$ \nl
Pr\,{\sc ii}& 0.75 & $-0.86$$\pm$0.23& 2 &$+0.28$ \nl
Nd\,{\sc ii}& 1.50 & $-0.68$$\pm$0.02& 2 &$+0.46$ \nl
\enddata
\end{deluxetable}

\clearpage

\begin{deluxetable}{lrr}
\tablecaption{Observed and Expected
Compositions}
\tablewidth{3in}
\tablehead{
\colhead{} &
 \colhead{[X/Fe]} &\colhead{[X/Fe]}\\
%\cline{2-3}
 \colhead{Species}& \colhead{Observed} &\colhead{Expected}}
\startdata
O & 0.6   & 0.7  \nl
Mg & 0.1   & 0.3  \nl
Si & 0.4    & 0.3  \nl
Ca & -0.1   & 0.3 \nl
Sc & 0.4  & 0.0  \nl
Ti & 0.2  & 0.3 \nl
V & 0.0   & 0.0  \nl
Cr & -0.1   & 0.0  \nl
Mn & -0.2   & -0.4  \nl
Co & 0.1  & -0.1  \nl
Ni & 0.0   & 0.0 \nl
Cu & -0.3   & -0.3  \nl
Y & 0.0    & 0.2  \nl
Zr & 0.2   & 0.0  \nl
Ce & -0.1    & -0.2 \nl
Pr & 0.3   & 0.2  \nl
Nd & 0.3  & 0.0 \nl
Sm & 0.3  & 0.1
\enddata
\end{deluxetable}